\documentstyle[preprint,aps,psfig]{revtex}
\draft
\def\la{\langle}
\def\ra{\rangle}
\def\beq{\begin{equation}}
\def\eeq{\end{equation}}
\def\be{\begin{eqnarray}}
\def\ee{\end{eqnarray}}

\def\k2av{\la k_T^2\ra}

\begin{document}
\preprint{KSUCNR-203-07}
\title{The charged-hadron/pion  ratio  at the Relativistic Heavy Ion Collider}
%\title{Cronin Effect and Hard Particle Ratio at RHIC}
\author{Xiaofei Zhang
and
George Fai}
\address{Center for Nuclear Research, Department of Physics,
Kent State University \\
Kent, Ohio 44242, USA}

\maketitle
%\vspace{-1.8in}
%{\hfill {KSUCNR-202-06}}

%\vspace*{1.8in}
\begin{abstract}
The hadron/pion ratio is calculated in 200 GeV $AuAu$ collisions at 
midrapidity, applying pQCD and non-universal
transverse-momentum broadening. 
Arguments are presented for such non-universality, and the idea
is implemented in a 
model, which   reproduces the main features of the centrality
dependence  of the hadron/pion ratio in  
$AuAu$ collisions. The model also  reasonably 
describes the qualitative difference
between the recently-measured $dAu$ nuclear enhancement factors 
for pions and charged hadrons.

\end{abstract}
\vspace{0.2in}


%\vspace{0.5in}
%\begin{multicols}{2}

Nuclear collisions are studied at unprecedented energies at 
the Relativistic Heavy Ion Collider (RHIC), revealing several
new phenomena embedded in a large amount of high-quality 
data, most aspects of which are consistent with general 
expectations\cite{QM2002}. A lot of attention is centered on 
final-state particles (secondaries) with high transverse momenta ($p_T$),  
where perturbative quantum chromodynamics (pQCD) should have good
predictive power. In this regime, a
suppression of total charged hadron and pion production
has been observed 
relative to a nucleon-nucleon ($NN$) reference in $AuAu$   
collisions\cite{Adams:2003kv,Adler:2003qi}. The suppression
can be described by pQCD calculations after incorporating final-state
partonic energy loss in dense 
matter\cite{Gyulassy:2000fs,Wang:2003mm,Baier:1998yf}, 
or using a model of    
initial-state gluon saturation\cite{Kharzeev:2002pc}. 
 The most recent
$dAu$ data\cite{usersmtg} provide important information  for 
understanding RHIC physics.

%The most recent
%$dAu$ data\cite{usersmtg} are expected to clarify the situation, and
%are thus of key importance in understanding RHIC physics.

Furthermore, the PHENIX collaboration
reports an anomalous enhancement of the high-$p_T$
proton-to-pion ($p/\pi$) ratio  
in $AuAu$ collisions at $\sqrt{s}=130$~GeV\cite{phen_ppi}
and $\sqrt{s}=200$~GeV\cite{Adler:2003kg}. 
At 200~GeV,
PHENIX also measured the ratio of charged hadrons 
($h^{\pm} = (h^+ + h^-)/2$) to neutral pions,
and found it enhanced in central collisions, presumably due to the 
enhancement of proton production. 
%An explanation for the $p/\pi$ enhancement was proposed recently, 
%combining pQCD with soft physics and jet quenching\cite{vitev_ppi}.
Different versions of   
coalescence (recombination)  of partons from the Qurak-Gluon Plasma   
were proposed to understand the 
enhanced  $p/\pi$ ratios\cite{coal}. Nevertheless,
the measured azimuthal correlations for high-$p_T$
charged particles\cite{Adler:2002tq} and 
the binary-collision scaling of proton production gleaned from
comparing different centralities\cite{Adler:2003kg}
indicate that the production from hard
scattering remains important.  Enhanced proton/pion and kaon/pion
ratios have also been observed in lower-energy pA collisions\cite{Cronin:zm}.
 Therefore, other
effects could also contribute to the enhanced particle ratio
at RHIC. In this paper, we will try to provide an alternative explanation
for the particle ratio data at RHIC based on pQCD.

%The particle ratios represent 
%one of the ``RHIC puzzles'' that deserve further study.

Another experimentally favored quantity, where differences 
between $h^{\pm}$ and $\pi$ secondaries manifest themselves, is the 
``nuclear modification factor'', $R_{AB}$. This ratio is designed 
to display the effects arising in an actual $AB$ nuclear collision, 
relative to a hypothetical collection of independent $NN$ collisions.
Earlier experimental information on $R_{AB}$ for 
$AuAu$ has recently been augmented by data obtained in the 
$dAu$ run to provide a crucial reference\cite{usersmtg}, and
the nuclear modification factor was also addressed 
theoretically\cite{lpbf,Vitev:2003xu,Wang:2003vy}.

Here we concentrate on the transverse-momentum and 
impact-parameter dependence of hadron/pion ratios at midrapidity and 
on the nuclear modification factor in $dAu$ at $\sqrt{s} = 200$~GeV, 
using leading-order pQCD. We observe that while 
pQCD is quite successful for  charged hadron and pion production 
at large $p_T$ in proton-proton ($pp$) collisions, proton production 
in $pp$ is not well understood using the language of pQCD. In fact,
pQCD underestimates the $p/\pi^+$ ratio by a factor of 3-10
in $pp$ collisions. This can be attributed to the limited 
information about a nonperturbative ingredient, the fragmentation 
function (FF), in the usual factorization-theorem based treatment of
particle production in pQCD (see $D(z,Q^2)$ in eq. (\ref{fullpipp})).
Most of the FF-s are extracted from $e^+e^-$ data, where
the most relevant large-z part of the FF-s is not well 
constrained\cite{Zhang:2002py}, and the information 
on proton FF-s is very limited, especially for gluon 
fragmentation functions. We are aware of only 
one set of proton FF-s from global fitting to date\cite{KKP}.
Compared to proton FF-s, FF-s for charged hadron and
pion production are better studied, supported by more  
experimental information\cite{Kretzer,BFGW}.
 At  the same time, the RHIC data  for the $p/\pi$  ratio are  only
available  for $p_T\le 5$ GeV, while   
$h^\pm/\pi$ ratios  are available for 
$p_T$   to  about 10 GeV. 
Therefore,
in this paper we focus on the $h^{\pm}/\pi^0$ 
ratio. Any insight gained is expected to also help in 
understanding the $p/\pi$ ratios. 
%It is important to realize
%that all present-day practical pQCD calculations have some
%uncertainty associated with truncations and nonperturbative
%ingredients. The essential aspects are expressed by robust 
%physical features, which are not too sensitive to modest
%variations. 

In an attempt to provide a more satisfactory description of available 
data and to mimic higher-twist contributions (of order $Q_v^2/p_T^2$,
where $Q_v$ is the appropriate virtuality\cite{Luo:fz}),
many pQCD calculations take direct account of the transverse momentum of
partons (``intrinsic $k_T$''). This can be accomplished via unintegrated 
parton distribution functions or, more phenomenologically, via a product  
assumption and a Gaussian transverse momentum distribution 
$ g({\bf k}_{T})$ (characterized by
the width $\langle k_T^2 \rangle $)\cite{Owens:1986mp,Zhang:2001ce}.
We apply the latter procedure in the present paper. Then, 
for $pp$ collisions, the usual convolution of the standard 
parton distribution functions (PDF-s) $f_{a}$, transverse momentum 
distributions $g({\bf k}_{T})$,
partonic cross sections $d\sigma/d{\hat t}$, and fragmentation 
functions (FF)  $D_{h/c}$ takes the form 
\begin{eqnarray}
\label{fullpipp}
&&E_{h}\frac{d\sigma_h^{pp}}{d^3p} =
        \sum_{abcd}  
        \int\!\!dx_{a,b} dz_c d^2k_{Ta,b} \
        g({\bf k}_{Ta}) g({\bf k}_{Tb}) \times
        \nonumber \\
        && f_{a}(x_a,Q^2) f_{b}(x_b,Q^2)\
             \frac{d\sigma}{d{\hat t}}
   \frac{D_{h/c}(z_c,{\widehat Q}^2)}{\pi z_c^2}{\hat s}
\delta({\hat s}+{\hat t}+{\hat u})  \, ,
\end{eqnarray}
where the partonic subprocesses $a+b \rightarrow c+d$ are summed over,
$x_a,x_b$ and $z_c$ are momentum fractions, and $\hat{s}, \hat{t}$ and
$\hat{u}$ denote the parton level Mandelstam variables.  
In this paper, we use CTEQ5L\cite{Lai:1999wy}
PDF-s and a set of FF-s\cite{Kretzer} which were determined relying also on
semi-inclusive data.

Figure 1 shows our results for $pp$ collisions. We apply fixed scales,
$Q = \widehat{Q} =p_T/2$, for both charged hadron and neutral pion production.
The top portions of the Figure compare the calculated 
charged hadron and pion spectra to
RHIC data\cite{Adams:2003kv,Adler:2003pb} for $p_T \geq 2$ GeV, 
where the pQCD calculation can be used. The data are
well described with $\k2av=1.8\pm 0.3$ GeV$^2$. 
The $h^{\pm}/\pi^0$ ratio is displayed in the bottom panel of 
Fig.~1. The ratio depends very weakly on $p_T$, and its value is
close to the peripheral $AuAu$ result ($\approx 1.6$)\cite{Adler:2003kg}. 
%The forthcoming $pp$ data will provide a precise test of the calculated
%$h^{\pm}/\pi^0$ ratio. 
The dependence of the cross sections 
on the scale is not weak,   limiting the predictive power of 
 leading order pQCD. Therefore we do not expect to fit the data 
in detail and   concerntrate on the main features. 
%However, 
 One  advantage of calculating ratios is
that the hadron-to-pion {\it ratio} is not sensitive to the scale chosen.
%(This is not unlike systematic errors canceling in certain 
%experimental ratios. The hadron-to-pion ratio is a good observable to
%test pQCD predictions.) 
The width $\k2av$ has a weak 
effect on the hadron to pion ratio as long as the same $\k2av$ value is used
for charged hadrons and for pions. (Later on in this paper we 
will introduce different values of $\k2av$ for different particle
species, but such differences should be small without medium enhancement,
with small effects on the $h^{\pm}/\pi^0$ ratio in $pp$ collisions.) 

It was observed in lower-energy proton-nucleus ($pA$) collisions
that the production of hard particles ($p_T \geq 2-3$~GeV) is 
enhanced more strongly than the naively expected scaling with 
$A$. This so-called Cronin effect\cite{Cronin:zm}
has been explained as initial state scattering or $k_T$ 
broadening\cite{Lev:1983hh}. 
The observed Cronin effect is not universal for different 
particle species.    
In $pA$ collisions at $\sqrt s= 38.8$~GeV, a stronger nuclear 
enhancement was seen in proton production than for pions\cite{Cronin:zm}.
We argue that the non-universality of the Cronin effect may be 
explained by the non-universality of the $k_T$ smearing. 
The latter can be understood 
in part by the large difference between pion and proton masses.
While the incoming parton does not ``know'' whether it will produce a pion or
a proton, it requires a larger $\hat{s}$ to produce a proton than a pion at
the same transverse momentum. This correlation connects initial-state 
broadening and final-state fragmentation. Furthermore,
as mentioned earlier, $k_T$ effects partly account for  higher-twist 
contributions of order $Q_v^2/p_T^2$. Assuming
larger $Q_v^2$ leads to larger power
corrections for protons or charged hadrons than for pions.

We make this observation the centerpiece of our description 
of the differences between the production of different hadrons
in the present study. Our strategy is to keep the other features 
of the model as simple as possible. We therefore write 
\begin{eqnarray}
\label{broad}
\k2av^h_{AB} & = & \k2av + c_h L_{AB}(b),  \nonumber \\ 
\k2av^\pi_{AB} & = & \k2av + c_\pi L_{AB}(b) \,\, ,
\end{eqnarray}
where $\k2av$ is the width of the transverse-momentum distribution 
in $pp$ collisions (here 1.8 GeV$^2$), and we only distinguish between
an average value for  charged hadron production, $\k2av^h$, 
and a value for pions $\k2av^\pi$, corresponding to the data we wish 
to consider. In both cases, we wrote the   
$k_T$-broadening as proportional to the effective length of the
medium along the path of the parton before the hard collisions, $L(b)$,
which depends on the impact parameter, $b$. 
In $pA$ collisions, $L_{pA}(b)$ is the average of the effective length,
\beq 
\label{effL}
L(z_A,{\bf b})=\int_{-\infty}^{z_A} dz'\rho_A(z',{\bf b})/\rho_0 \,\, ,
\eeq
with $(z_A,{\bf b})$ representing the point of the hard collision
and $\rho_0$ the average density of the target, 
over the Glauber nuclear thickness function of the nucleus. 
The quantity~(\ref{effL}) integrated over $\bf b$ 
is proportional to $A^{1/3}$.
In $AB$  collisions, the effective length for a hard parton from nucleus $A$
passing through nucleus $B$,  
$L_{AB}(b)$, is the average of $L(z_B,{\bf s}_B)$ over the nuclear thickness
function for the collision of nuclei $A$ and $B$, at the given value 
of ${\bf b}$, which is proportional to  the probability of 
hard  collisions at a certain vector impact parameter {\bf b}.
In this paper we use Woods-Saxon nuclear density profiles, with the 
parameter values for $Au$ taken from Ref. \cite{W-S}. The effective length 
can also be written in terms of the number of collisions suffered by
the incoming nucleon, $\nu(b)$\cite{Zhang:2001ce}. 
For our schematic average purposes we find it more appropriate 
to use the effective length $L(b)$.   

Here we focus on the difference between the coefficients $c_h$ and $c_\pi$.
If the proton contribution to  charged hadrons varied with impact 
parameter, then the composition-dependent
$c_h$ should also be expected to depend on $b$ or  $p_T$. 
However, we will show 
that the data can be fitted well by $b$
and  $p_T$ independent coefficients.
This may be understood in terms of a $p_T$-integrated proton contribution
to the  charged hadron yield, which does not strongly depend on $b$
or $p_T$.
Thus, we neglect any potential $b$ and  $p_T$  dependence of $c_h$.
We expect $c_h$ to be larger than $c_\pi$.  
To calculate hadron and pion cross sections separately, one needs to 
account for additional nuclear effects like e.g. 
shadowing and the suppression mentioned in the introduction.
Testing various shadowing parameterizations, we found that
the shadowing effects are not important to the $h^\pm/\pi^0$ ratio. 
Similarly, if the suppression factors are not too different, 
we expect an approximate cancellation in cross section ratios. 
We make this assumption in the following calculation.

%One could use the $dAu$ data at $\sqrt{s} = 200$~GeV
%to fix the parameters in eqs. (\ref{broad}), and apply the results 
%to $AuAu$ reactions. (We consider the loosely-bound deuteron as a 
%superposition of a proton and a neutron.) Alternatively,
%the hadron-to-pion data in $AuAu$ collisions can be
%used to fix the parameters and to predict the $dAu$ results. 
%The advantage of this procedure is that the $h^{\pm}/\pi^0$ {\it ratios} 
%are less sensitive to shadowing and suppression effects\cite{note}.

Figure 2 shows the hadron-to-neutral-pion ratio for different 
centralities in $AuAu$ collisions at $\sqrt{s} = 200$~GeV  
with $c_\pi=0.13\pm 0.04$ GeV$^2$/fm 
and $c_h=0.45\pm 0.08$ GeV$^2$/fm. 
The agreement appears to be satisfactory,
 and is similarly good for 
other centrality bins (not shown). 
  As we expected   $c_h>c_\pi$.
In other words,  
%with larger mass 
the effective $\hat s$ (the
energy involved in the partonic cross section) is larger 
at the same $p_T$ for the average hadrons than for pions.
Larger $\hat s$ leads to more room for a dynamical intrinsic $k_T$,
just like in the Drell-Yan case, where the larger $Q_v^2$ of the lepton
pair leads to a larger $k_T^2$~\cite{QZ_resum}.  
Therefore we find $c_h>c_\pi$ natural, while we do not have
a quantitative understanding of the fact that $c_h\approx 3.5 c_\pi$.
We noticed that  the trend of   the $h^\pm/\pi$  ratio at large $p_T$ 
for     the $60-70\%$ bin  
is somewhat different from the trend of other bins.  
We hope future data with smaller error
bars will clarify this point.

As the collisions become more central, the effective 
length increases, and the average 
transverse momentum associated with charged hadrons will
broaden more than that of pions. This leads to the 
enhancement of the charged-hadron-to-pion ratio.
More study is needed to further understand the origin of the 
difference between the coefficients $c_h$ and $c_\pi$.

Now, let us apply the above parameters to  calculate the $R_{dAu}$
nuclear enhancement factors for  charged hadron and neutral pion
production. It is believed that final state effects are 
much less important in $dAu$ than in $AuAu$, and {initial} $k_T$-broadening may play 
a key role. We carry out the calculation with both 
unmodified PDF-s\cite{Lai:1999wy} and nuclear PDF-s which incorporate 
``shadowing''. Here we use the EKS parameterization\cite{Eskola:1998df}.   
The results are displayed in Fig.~3, together with 
the recently released PHENIX data\cite{usersmtg}. 
One important feature  at 2 GeV $\le p_T<6$ GeV
is that the $R_{dAu}$ nuclear modification factors 
for  charge hadrons (solid line: without shadowing; dashed: with 
EKS shadowing) and neutral pions (dotted: without shadowing; dot-dashed: 
with EKS shadowing) are different. 
%The shape of the $h^{\pm}$ curves
%reproduces the main features of the data above 2~GeV. 
The EKS shadowing increases $R_{dAu}$ somewhat, since we 
are in the anti-shadowing region. The nuclear modification factors for pions
are generally much smaller than the ones for $h^{\pm}$, in agreement with
the data\cite{usersmtg}. This is because, 
as observed e.g. in Ref. \cite{lpbf}, the proton-proton $\k2av$ determines
the position of the Cronin peak, 
while the coefficient $c$ in eq. (\ref{broad}) regulates the height of the 
peak. The non-universal $k_T$-broadening
may further  contribute to  the different behavior of 
the nuclear modification factor for $h^{\pm}$ and for pions in $AuAu$ 
collisions.
%If suppression effects like jet quenching do not strongly 
%depend on particle species as we assumed, the calculated results 
%presented here will not change much when jet quenching is taken into
%account. 
It will however be interesting  to carry out a 
less schematic calculation to understand
the different shapes of $R_{AB}$ for  charged hadrons
and neutral pions, respectively.   
 In addition, data for the nuclear modification factors are becoming available
species by species. A less schematic calculation should also address them to
 distinguish mass effects and effects of particle species\cite{phi}.

In summary, we explored the consequences of different Cronin enhancement 
coefficients in the widths of the transverse-momentum distributions of
different secondaries in this paper. We argued for this non-universality 
on the basis of lower-energy observations and on theoretical grounds, 
related to the higher-twist structure of
pQCD calculations. It was found that 
%an explanation of the hadron/pion puzzle, or more in detail, 
 the main features of
the 
transverse-momentum and impact-parameter dependence of hadron/pion
ratios at $\sqrt{s} = 200$ GeV can be obtained  in a 
simplistic model if the non-universality is incorporated. The
 different nuclear modification factors recently measured in $dAu$ collisions are  also displayed by the model. 
Enhancement of the charged-hadron/pion ratio in $dAu$ compared
to $pp$ collisions is a direct prediction of this model, which 
could be tested in the near future.
The understanding of further intriguing features
of nuclear modification factors in $AuAu$ collisions is left for future work.
   
We are grateful to D. D'Enterria, J. Jia, D. Keane, and J. W. Qiu for 
stimulating discussions. X. F. Zhang acknowledges the hospitality of the
Institute for Nuclear Theory, where part of this work was carried out.
This work
was partially supported by the U.S. DOE under DE-FG02-86ER-40251.

%\vspace*{2 mm}

%\end{multicols}

\begin{figure}
\begin{center}
\psfig{figure=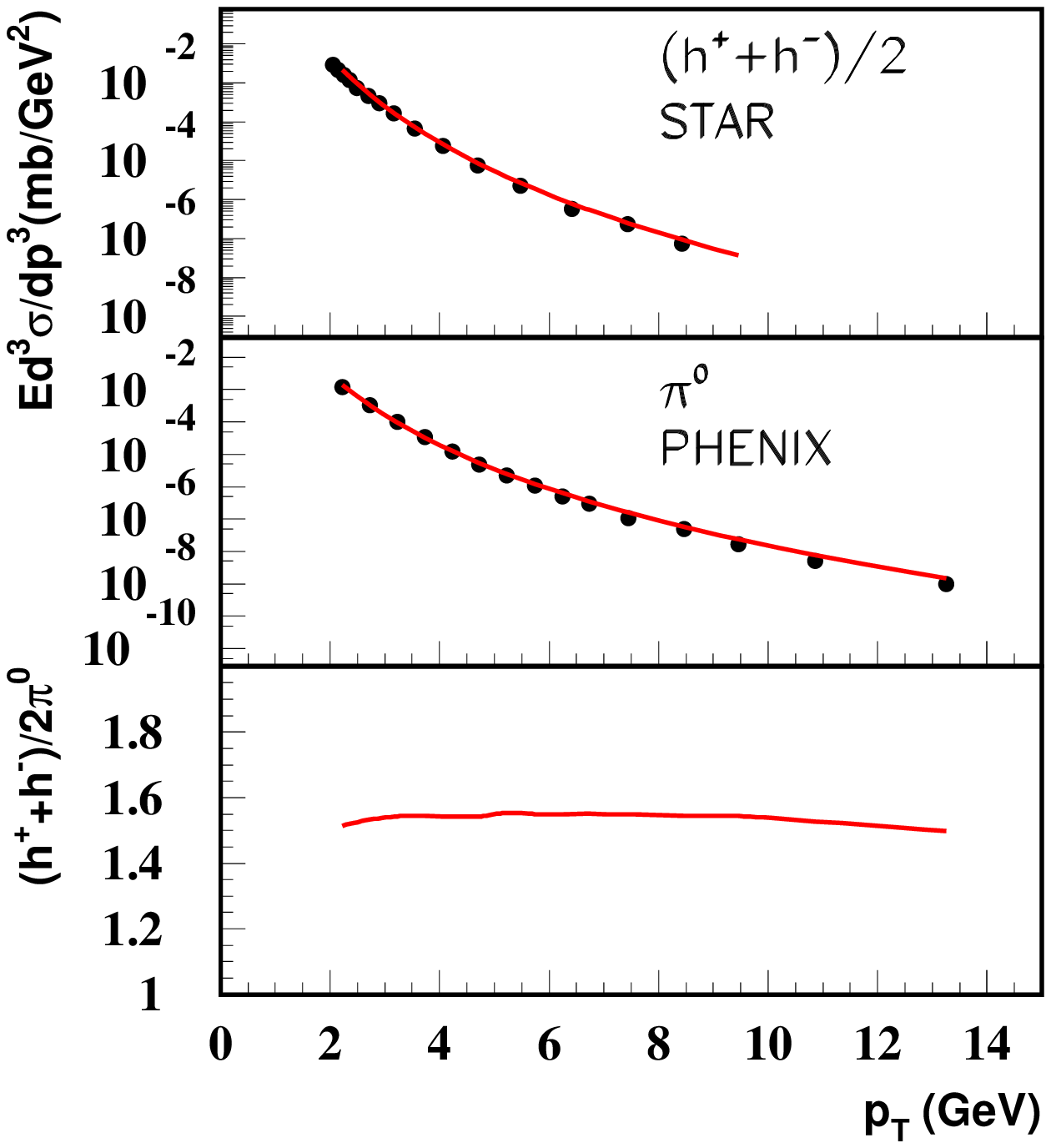,width=4.5in}
\end{center}
%\vspace{-0.3in}
\caption{ 
Invariant cross section of charged hadron (top) 
and pion (middle)  production  in $pp$ collisions
at $\sqrt{s} = 200$~GeV. The data points are from
STAR\protect\cite{Adams:2003kv} and  
PHENIX\protect\cite{Adler:2003pb}; the solid 
lines represent the leading-order pQCD calculation with $\k2av=1.8$~GeV$^2$.
Bottom: hadron/pion ratio calculated 
with $\k2av=1.8$ GeV$^2$ for both, pion and charged hadron
production.}
\end{figure}

\begin{figure}
\begin{center}
\psfig{figure=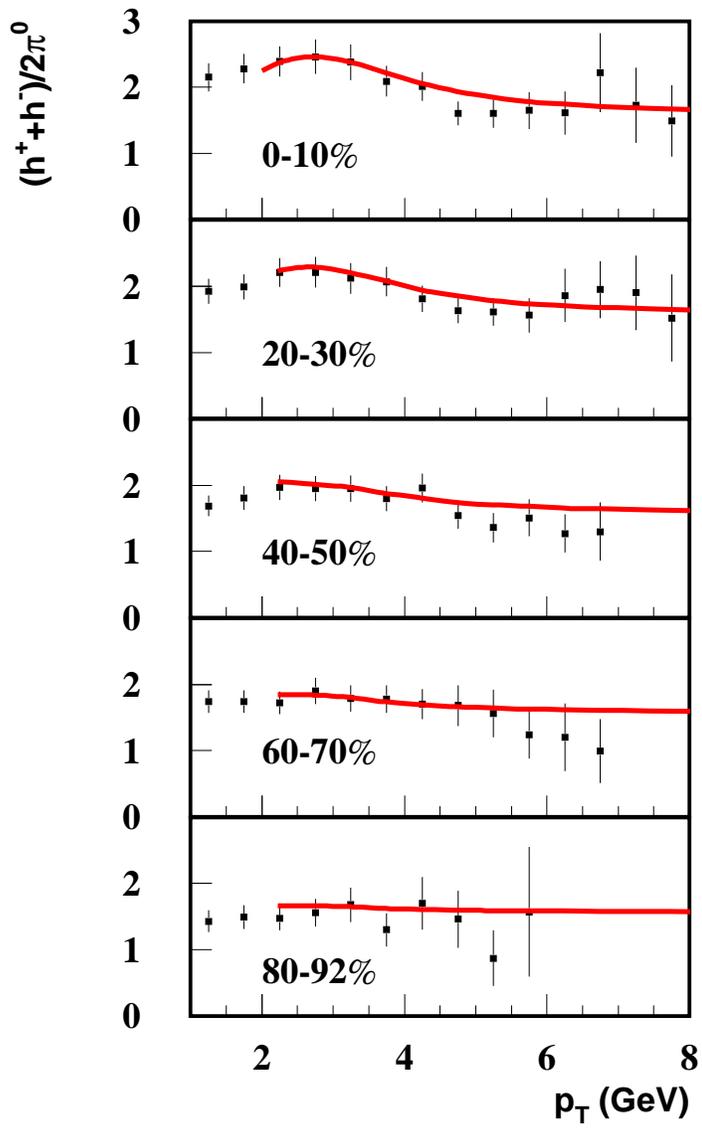,width=4.5in}
\end{center}
%\vspace{0.15in}
\caption{ 
Hadron-to-pion ratios 
at different centralities. From top to bottom: 
0-10\%;  20-30\%  ...
(solid lines);
data are from PHENIX(normalization errors are not included)
\protect\cite{Adler:2003kg}.}
\end{figure}

\begin{figure}
\begin{center}
\psfig{figure=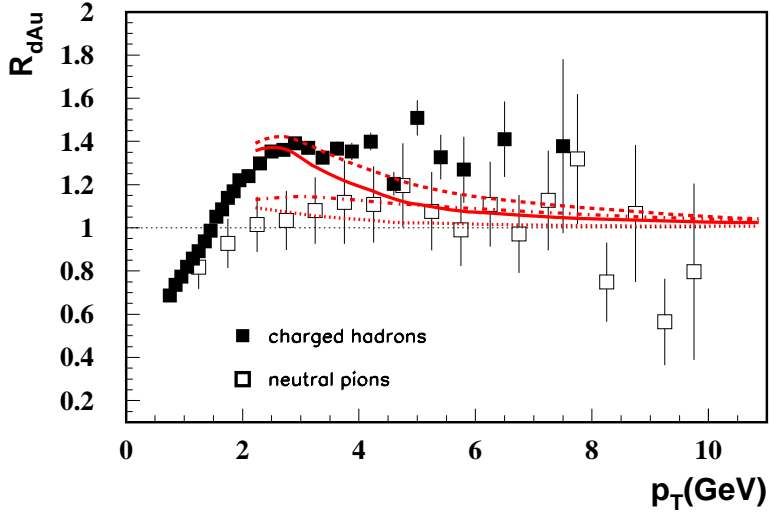,width=4.5in}
\end{center}
%\vspace{-0.3in}
\caption{ 
The nuclear modification factor $R_{dAu}$ for $h^{\pm}$
(solid line: without shadowing; dashed: with 
EKS shadowing) and $\pi^0$ (dotted: without shadowing; dot-dashed: 
with EKS shadowing. Data points are from PHENIX $dAu$ data
\protect\cite{usersmtg}.}
\end{figure}

\end{document}